\documentclass[useAMS,usenatbib]{mn2e}
\usepackage{times}
\usepackage{arydshln}

\usepackage{graphicx}	
\usepackage{amsmath}	
\usepackage{amssymb}	

\usepackage{rotating}

\def\nustar{\textit{NuSTAR}}
\def\xmm{\textit{XMM-Newton}}
\def\xspec{\textsc{xspec}}
\input{epsf}
\usepackage[caption=true]{subfig}    

\title[Iron line modulation in H 1743-322]{A quasi-periodic modulation
  of the iron line centroid energy in the black hole binary H 1743-322}

\author[A. Ingram et al]{Adam
Ingram,$^{1}\thanks{E-mail:a.r.ingram@uva.nl}$
Michiel van der Klis,$^1$
Matthew Middleton,$^2$
Chris Done,$^3$
\newauthor
Diego Altamirano,$^4$
Lucy Heil,$^1$
Phil Uttley$^1$
\& Magnus Axelsson$^5$\\
$^1$Anton Pannekoek Institute for Astronomy, University of Amsterdam,
Science Park 904, 1098 XH, Amsterdam, The Netherlands.\\
$^2$Institute of Astronomy, Cambridge University, Madingley Road, CB3 0HA, Cambridge, UK \\
$^3$Center for Extragalactic Astronomy, Department of Physics, University of Durham, South Road, Durham, DH1 3LE, UK \\
$^4$Department of Physics \& Astronomy, University of Southampton, Southampton, Hampshire SO17 1BJ, UK \\
$^5$Department of Physics, Tokyo Metropolitan University, Minami Osawa
1-1, Hachioji, Tokyo 192-0397, Japan }

\begin{document}

\date{Accepted 2016 May 19. Received 2016 May 19; in original form 2016 April 15}

\pagerange{\pageref{firstpage}--\pageref{lastpage}} \pubyear{2016}

\maketitle

\label{firstpage}

\begin{abstract} 
\noindent Accreting stellar-mass black holes often show a `Type-C'
quasi-periodic oscillation (QPO) in their X-ray flux, and an iron
emission line in their X-ray spectrum. The iron line is generated
through continuum photons reflecting off the accretion disk, and its
shape is distorted by relativistic motion of the orbiting plasma and
the gravitational pull of the black hole. The physical origin of the
QPO has long been debated, but is often attributed to Lense-Thirring 
precession, a General Relativistic effect causing the inner flow to
precess as the spinning black hole twists up the surrounding
space-time. This predicts a characteristic rocking of the iron line
between red and blue shift as the receding and approaching sides of
the disk are respectively illuminated. Here we report on \xmm~and
\nustar~observations of the black hole binary H 1743-322 in which the
line energy varies systematically over the $\sim 4$ s QPO cycle
($3.70\sigma$ significance), as predicted. This provides strong
evidence that the QPO is produced by Lense-Thirring precession,
constituting the first detection of this effect in the strong
gravitation regime. There are however elements of our results harder
to explain, with one section of data behaving differently to all the
others. Our result  enables the future application of tomographic
techniques to map the inner regions of black hole accretion disks.
\end{abstract}


\section{Introduction}
\label{sec:intro}

Accreting stellar-mass black holes routinely exhibit `Type-C' low
frequency quasi-periodic oscillations (QPOs) in their X-ray flux, with
a frequency that evolves from $\sim 0.1 - 30$ Hz as the X-ray spectrum
transitions from the power-law dominated hard state to the thermal
disk dominated soft state (e.g. \citealt{Wijnands1999};
\citealt{VDK2006}). The thermal disk component is well understood as
originating from a geometrically thin, optically thick accretion disk
(\citealt{Shakura1973}; \citealt{Novikov1973})
and the power-law emission, which displays breaks at low and high
energy, is produced via Compton up-scattering of seed photons by a
cloud of hot electrons located close to the black hole
(\citealt{Thorne1975}; \citealt{Sunyaev1979}). The low and high energy
breaks are associated respectively with the seed photon temperature
and the electron temperature. The exact geometry
of this electron cloud is uncertain, and is probably changing through
this transition. In the \textit{truncated disk
  model} (\citealt{Ichimaru1977}; \citealt{Esin1997};
\citealt{Poutanen1997}), the disk truncates in the hard state at some
radius larger than the innermost stable circular orbit (ISCO), with
the inner regions forming a large scale height, hot accretion flow
(hereafter the \textit{inner flow}) which emits the Comptonised
spectrum. The Comptonised spectrum becomes softer and the disk
component becomes more prominent in the spectrum as the truncation
radius moves inwards (e.g. \citealt{Done2007}). Alternatively, the hot 
electrons may be located in a corona above the disk
(e.g. \citealt{Galeev1979}; \citealt{Haardt1991}) or
at the base of a jet (e.g. \citealt{Markoff2005}), or perhaps some
combination of these alternatives. The X-ray spectrum also
displays reflection features, formed by Comptonised photons being
scattered back into the line of sight by the disk. The most prominent features of
the reflection spectrum are the iron K$_\alpha$ line at $\sim 6.4$
keV, formed via fluorescence, and the reflection hump peaking at $\sim
30$ keV, formed via inelastic scattering from free electrons
(e.g. \citealt{Ross2005}; \citealt{Garcia2013}). The shape of the
reflection spectrum, and in particular the iron line which is narrow
in the rest frame, is distorted by orbital motion of the disk material
and gravitational redshift (\citealt{Fabian1989}).

The QPO arises from the immediate vicinity of the black
hole\footnote{the light crossing timescale puts a hard upper limit of
  $\sim 300~R_g$ (where $R_g=GM/c^2$), but the true size scale is
  likely $\lesssim 60~R_g$ (e.g. \citealt{Axelsson2013}).}, but its
physical origin has long been debated. Suggested QPO models in the
literature generally consider either some instability in the accretion
flow, or a geometric oscillation. Instability models consider, for
example, oscillations in mass accretion rate or pressure
(e.g. \citealt{Tagger1999}; \citealt{Cabanac2010}) or standing shocks
in the disk (e.g. \citealt{Chakrabarti1993}). Geometric models mostly
consider relativistic precession. Due to the frame dragging effect, a
spinning black hole drags the surrounding spacetime around with it,
inducing Lense-Thirring precession in the orbits of particles out of
the equatorial plane (\citealt{Lense1918}). \cite{Stella1998} and
\cite{Stella1999} were the first to suggest that low frequency QPOs
could be driven by Lense-Thirring precession, noting that the
expected precession frequency of a test mass at the truncation radius
is commensurate with the QPO frequency. \cite*{Schnittman2006}
considered a precessing ring in the disk, and corrugation modes in the
disk caused by the frame dragging effect have also been studied
(e.g. \citealt{Wagoner2001}). \cite*{Ingram2009} suggested that the
entire inner flow precesses whilst the disk remains stationary,
motivated by the simulations of \cite{Fragile2007}. This model
explains why the QPO modulates the Comptonised emission much more than
the disk emission, and predicts that the QPO should be stronger in
more highly inclined sources as observed (\citealt*{Schnittman2006};
\citealt*{Heil2015}, \citealt{Motta2015}). It also makes a distinctive
prediction: as the inner flow precesses, it will illuminate different
azimuths of the disk such that an inclined observer sees a blue/red
shifted iron line when the approaching/receding sides of the disk are
illuminated (\citealt{Ingram2012a}). The precession model therefore
predicts that the line energy changes systematically with QPO
phase. This is a difficult effect to measure, since phase-resolving
the QPO poses a technical challenge. \cite{Miller2005} used a
simple flux selection to obtain suggestive but inconclusive
results for GRS 1915+105. \citet[hereafter~IK15]{Ingram2015} developed a more
sophisticated technique to discover spectral pivoting and a
  modulation in the iron line flux in the same source, but data
  quality prevented unambiguous measurement of a line energy
  modulation. Recently, \cite{Stevens2016} developed a similarly
  sophisticated QPO phase-resolving technique, which involves
  cross-correlating each energy channel with a reference band. Using
  this technique, they found a modulation in the disk temperature of
  GX 339-4, interpreted as reprocessed radiation from a precessing
  inner flow or jet. However, they too lacked the data quality to measure a
  line energy modulation.

In this paper, we further develop the QPO phase-resolving method of
IK15, conducting fitting in the Fourier domain rather than the time
domain so that the error bars are independent. We use this method to
analyse a long exposure observation of the black hole binary H
1743-322 in the hard state. We summarise the observations in Section
\ref{sec:data}, describe our phase-resolving method in Section
\ref{sec:method} and present the results of fitting a phenomenological
model to the phase-resolved spectra in Section \ref{sec:results}. We
discuss our findings in Section \ref{sec:discussion} and outline our
conclusions in Section \ref{sec:conclusions}.

\section{Observations}
\label{sec:data}

\begin{figure}
 \includegraphics[height= 8.5 cm,width=4cm,trim=0.0cm 0.0cm 0.0cm
0.0cm,clip=false,angle=-90]{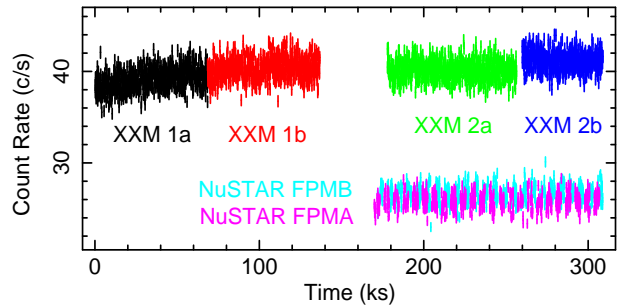}
\vspace{0mm}
 \caption{Long term light curve summarising the \xmm~and \nustar~observations analysed in this
   paper. Throughout, the data are referred to as labelled in this
   plot. The rise in count rate from \xmm~orbit 2a to
   2b is due to a small change in the instrument setup (from pn thick
   to pn thin) as a new PI took over the observation.}
 \label{fig:lc}
\end{figure}

The \textit{X-ray Multi-Mirror Mission} (\xmm; \citealt{Jansen2001})
observed H 1743-322 for two full orbits of the satellite around the
Earth in late September 2014. The first orbit  (obs ID 0724400501)
lasted from $\sim$18:45 on 21$^{\rm st}$ September until 
$\sim$08:45 on 23$^{\rm rd}$ September. The second orbit lasted from
$\sim$20:10 on 23$^{\rm rd}$ September until $\sim$08:35 on 25$^{\rm
  th}$ September. The second orbit is split into two obs IDs, with the
first $\sim 70$ ks classified as 0724401901 and the final $\sim 50$
ks, which had a different PI, as 0740980201. In this paper, we split
up each orbit into two separate observations to allow for potential
evolution of spectral and timing properties over such long exposures
(and also the small change in instrumental setup as the PI
changed). Hereafter, we refer to these four \xmm~observations as
orbits 1a, 1b, 2a and 2b. The \textit{Nuclear SpecTroscopic ARray} (\nustar;
\citealt{Harrison2013}) observed the source from $\sim$18:20 on
$23^{\rm rd}$ September until $\sim$08:50 on $25^{\rm th}$ September
(obs ID 80001044004). Figure \ref{fig:lc} shows long term $4-10$ keV
light curves for all exposures and illustrates our naming convention
for the \xmm~data. Spectral and timing analyses of the
  \xmm~data have been previously presented by \cite{Stiele2016} and
  \cite{DeMarco2016}, whereas the \nustar~data are reported on here
  for the first time.

\subsection{Data reduction}
\label{sec:datred}

\subsubsection{\xmm}

We used the \xmm~Science Analysis Software (SAS)
version 14.0 to reduce data from the EPIC-pn (European Photon Imaging
Camera) in timing mode. We generated calibrated and concatenated event
lists using \textsc{epproc} with the default settings for timing mode
as of SAS v14.0 (runepreject=yes withxrlcorrection=yes runepfast=no
withrdpha=yes). We extracted all products from a region
$32\leq$RAWX$<44$, RAWY $\geq 23$ and use only single and double 
events (PATTERN $\leq 4$), whilst ignoring bad pixels (FLAG==0). We
generated response and ancillary files using \textsc{rmfgen} and
\textsc{arfgen}, and rebinned all spectra to have at least 20 counts
per channel using \textsc{specgroup}. We extracted background spectra
from the region $3\leq$RAWX$\leq 5$, RAWY $\geq 23$ and find that the
source contributes $98.5 \%$ of the total counts (this number is
likely even higher in reality, since source counts can contaminate the
background spectrum in timing mode: \citealt{Done2010}). Since the
source dominates, we did not perform a background subtraction when
extracting light curves. Inspection of the long term $10-12$ keV light
curve reveals that none of the exposure is affected by proton flares.

We extract light curves in 20 energy bands. We focus our
phase-resolved analysis on the $4-10$ keV region, so extract one broad
light curve for energies $< 4$ keV and one broad light curve for
energies $> 10$ keV (both of which will be ignored for the analysis),
leaving 18 high signal-to-noise-channels in the region of
interest. These channels are broad enough to achieve good statistics,
and are trivially broader than the FWHM of the instrument response. We
used the \textsc{ftool} \textsc{rbnrmf} in order to re-bin the
spectral response file into these 20 energy bands.

\subsubsection{\nustar}

We used the \nustar~analysis software, NuSTARDS v1.4.1. We
extracted products from the cleaned event list with the \textsc{ftool}
\textsc{nuproducts}, using a 120'' circular source extraction region
and a 90'' circular background extraction region taken from an area
not contaminated by source counts. We find that the source contributes
$99.7 \%$ of the total counts, and consequently we did not perform a
background subtraction when extracting light curves. The background is
negligible up to $\sim 50$ keV, above which it dilutes the rms and
phase lags by a small amount. We extract light curves in 19 energy
bands. We concentrate on the energy range $4-60$ keV, and so bin into
2 broad channels for energies $< 4$ keV and 1 broad channel $> 60$ keV
(with these three channels to be ignored in the analysis), leaving 16
high signal-to-noise channels in the range of interest. As for \xmm,
we re-binned the spectral response file using \textsc{rbnrmf}.

\subsection{Power spectra}
\label{sec:tim}

\begin{figure}
 \includegraphics[height= 7.5cm,width=8.5cm,trim=0.8cm 0.0cm 0.0cm
0.0cm,clip=true]{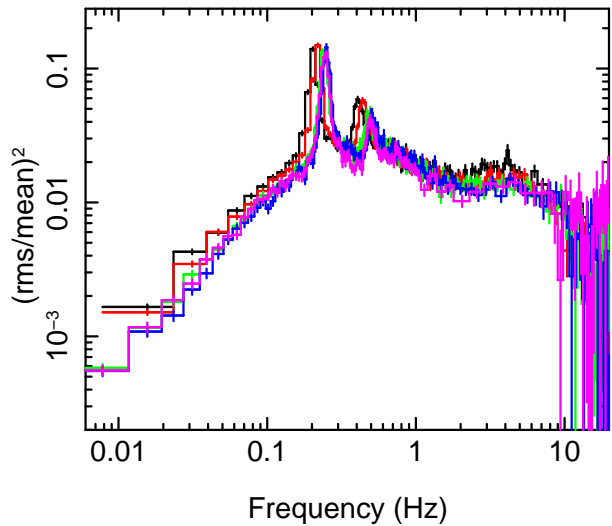}
\vspace{-8mm}
 \caption{
$4-10$ keV Power spectrum for \xmm~orbits 1a (black), 1b (red), 2a
(green) and 2b (blue), and $4-10$ keV co-spectrum between the \nustar~FPMA
and FPMB (magenta).For all datasets we see a strong Type-C QPO with
two clearly detected harmonics. The QPO frequency increased from $\sim
0.2$ Hz to $\sim 0.25$ Hz over the $\sim 300$ ks elapsed from the
start of orbit 1a to the end of orbit 2b. Error bars are 1 $\sigma$.}
 \label{fig:psd}
\end{figure}

Fig. \ref{fig:psd} shows 4-10 keV power spectra calculated for
\xmm~orbits 1a (black), 1b (red), 2a (green) and 2b
(blue) and NuSTAR (magenta). The \xmm~power spectra are
calculated in the standard way, with a constant Poisson noise level
subtracted (\citealt{VdK1989}; \citealt{Uttley2014}). For \nustar~we
instead calculate the co-spectrum between the two (independent) Focal
Plane Modules, FPMA and FPMB (\citealt{Bachetti2015}), since the
\textit{NuSTAR} dead time of $\tau_d \approx 2.5$ ms imprints
instrumental features on the Poisson noise in a power spectrum
calculated in the standard way. The co-spectrum is the real part of
the cross-spectrum and includes no Poisson noise contribution. We also
correct for the suppression of variability caused by the \nustar~dead
time using the simple formula (\citealt{Bachetti2015})
\begin{equation}
\frac{{\rm rms}_{\rm det}}{{\rm rms}_{\rm in}} \approx
\frac{1}{1+\tau_d r_{\rm in} } = \frac{r_{\rm det}}{r_{\rm in}},
\label{eqn:deadc}
\end{equation}
where $r_{\rm det}$ and $r_{\rm in}$ are respectively the detected
and intrinsic count rates. For this observation, the ratio of detected
to intrinsic variability is ${\rm rms}_{\rm det} / {\rm rms}_{\rm in}
= 0.8462$ (recorded in the NuSTAR spectral files as the keyword
`DEADC'). The power spectra in Fig. \ref{fig:psd} are normalised such
that the integral of the power spectrum over a given frequency range
gives the variance of the corresponding time series over that range,
and are plotted in units of frequency $\times$ power.

All power spectra display QPOs with a strong fundamental (first
harmonic) and overtone (second harmonic) evidenced by two large,
harmonically related peaks. We see that the QPO fundamental frequency
evolved from $\sim 0.205$ Hz to $\sim 0.25$ Hz over the $\sim 300$ ks
duration of the two \xmm~orbits. We also see that the 4-10 keV (dead
time corrected) \nustar~co-spectrum agrees very well with the
simultaneous \xmm~orbit 2 data for the same energy band. For our
analysis, we treat each of the five datasets shown in
Fig. \ref{fig:psd} separately to allow for the evolution in source
properties over such a long exposure, and also to allow for the
different responses of the two instruments, and the small change in the
\xmm~instrumental setup during orbit 2.

\subsection{Energy spectra}
\label{sec:spec}

As a preliminary analysis, we jointly fit the spectra of both
\xmm~orbit 2a and the simultaneous (FPMA) \nustar~observation with a
simple absorbed power-law plus Gaussian iron line model, considering
only $4-10$ keV for both. Throughout this paper, we account for
interstellar absorption using the model \textsc{tbabs}, with hydrogen
column density $N_H = 1.35 \times 10^{22} {\rm cm}^{-2}$ and the
relative abundances of \cite{wilms2000}. We use
\xspec~v12.8.2 for all spectral fitting (\citealt{Arnaud1996}). We
achieve a best fit with reduced $\chi^2=551.14/529=1.04$, without
applying any systematic error. There is no evidence for direct disk
emission in the $> 4$ keV bandpass, and the \nustar~spectrum above
$10$ keV reveals a reflection hump. In this paper, we focus on
phenomenological modelling of the $4-10$ keV region for our
QPO phase-resolved analysis, modelling continuum and iron line with a
power-law and Gaussian respectively. We consider this bandpass because
it is shared between \xmm~ and \nustar, it is above the energies for
which direct disk emission is relevant and below energies for which
the reflection hump is important. Clearly, a Gaussian function is not
a physical model for the iron line, but we wish to characterise the
QPO phase dependence of the iron line profile without making physical
assumptions. We will focus on physical spectral modelling in a future
paper.

We find a discrepancy in the power-law index measured for these two
spectra ($1.286\pm 0.003$ for \xmm~and $1.509\pm0.004$ for
\nustar). The Gaussian representing the iron line has a larger
equivalent width in the \nustar~spectral fit ($\sim 65$ eV) than in
the \xmm~data ($\sim 47$ eV), and lower centroid energy in the
\nustar~data ($\sim 6.41$ keV) than in the \xmm~data ($\sim 6.61$
keV), but the line width is consistent. This cross-calibration
discrepancy poses a problem for time-averaged spectral
analysis. However, our analysis is differential: it focuses on the
variation of spectral parameters with QPO phase, and is therefore far
more robust to cross-calibration issues. We demonstrate in the
following two sections that the variability properties are consistent
between the two observatories, and that the differential variation in
each of the spectral parameters with QPO phase is consistent. For our
phase-resolved spectral analysis, we allow the time averaged power-law
index and line energy to be different between the two observatories,
but tie their \textit{differential} properties between the two
observatories.

\section{Phase-resolving method}
\label{sec:method} 

We use the phase-resolving method of
IK15, with some small changes designed to
increase signal-to-noise and circumvent the \textit{NuSTAR} dead
time, and also some more significant changes to allow us to reliably
calculate statistical significances for our fits. The essence of the
IK15 phase-resolving method is to measure the Fourier transform (FT)
of the QPO as a function of energy $E$, for each harmonic for which
this is possible. For the $j^{\rm th}$ harmonic, this can be written as
\begin{equation}
W_j (E) = \mu(E) \sigma_j(E) e^{ i \Phi_j(E) },
\label{eqn:qpoft}
\end{equation}
where $\mu(E)$ is the mean count rate, $\sigma_j(E)$ the fractional
rms in the $j^{\rm th}$ QPO harmonic and $\Phi_j(E)$ is referred to in
IK15 as the \textit{phase offset} of the $j^{\rm th}$ harmonic, all as
a function of energy. It is clear from Fig. \ref{fig:psd} that the QPO
for the observations considered here has two strong harmonics,
therefore we calculate the QPO FT for $j=1$ and $j=2$. We must also
consider the case of $j=0$; i.e. the DC component (standing for direct
current). This is simply the mean count rate, such that $W_0(E) =
\mu(E)$.

As for the phase offsets, $\Phi_j(E)$, we can calculate the
cross-spectrum between each energy channel and a reference band in
order to measure the phase lag for each harmonic, $\Delta_j(E)$, as a
function of energy. That is, we can measure by how many radians the
$j^{\rm th}$ harmonic of each energy channel lags the $j^{\rm th}$
harmonic of the reference band. What we can \textit{not} measure using
the cross-spectrum is the phase difference between the harmonics. By
measuring this phase difference, we can calculate the phase-offsets of
the first two harmonics using the formulae
\begin{eqnarray}
\Phi_1(E) &=& \Phi_1 + \Delta_1(E) \nonumber \\
\Phi_2(E) &=& 2 [ \Phi_1 + \psi ] + \Delta_2(E).
\label{eqn:Phi}
\end{eqnarray}
Here, $\psi$ is the phase difference between the two harmonics in the
reference band and $\Phi_1$ is the arbitrary reference phase of the
first harmonic, which we set to $\Phi_1=\pi/2$ following IK15. Note
that there is a version of the above formula in IK15 (equation 8 in
that paper), which differs slightly from equation \ref{eqn:Phi}
presented here. The version presented here is correct and the mistake
is in IK15. Note that we only need to measure the phase difference
between the harmonics in one band (is it obviously advantageous to
measure this for the reference band which has far more photons than
the individual channels). The phase difference between the harmonics
as a function of energy is given by $\psi(E) = \psi - \Delta_1(E) +
\Delta_2(E)/2$. We stress that the use of a broad reference band does
\textit{not} smear out the data in some way, as is a common
misconception. For unity coherence, changing the reference band
affects \textit{only} the constant offset of the lag spectrum and also
the signal-to-noise (see e.g. \citealt{Uttley2014}).

The method of IK15 involves taking the inverse FT of
equation \ref{eqn:qpoft} to give an estimate of the QPO waveform in
each energy channel. This method is intuitive, since it gives a way of
estimating the spectrum as a function of QPO phase. The inverse FT,
however, introduces correlations in the errors between different QPO
phases. Here, we first summarise the time domain approach and then
describe a new Fourier domain approach that circumvents the problem of
correlated error bars associated with the time domain method.

\subsection{Phase-resolved spectra in the time domain}

We can inverse FT equation \ref{eqn:qpoft} to estimate the QPO
waveform for each energy band
\begin{equation}
w(E,\gamma)= \mu(E) \left\{  1 + \sqrt{2} \sum_{j=1}^2 \sigma_j(E)
  \cos[ j\gamma - \Phi_j(E) ]   \right\},
\label{eqn:waveform}
\end{equation}
where $\gamma$ is QPO phase. Plotting this instead as count rate
versus photon energy for a given QPO phase gives phase-resolved
spectra. We describe in the following subsections how we measure
$\mu(E)$, $\sigma_j(E)$ and $\Delta_j(E)$, focusing mainly on the
modifications we have made to the IK15 method in order to maximise
signal-to-noise and correct for the \nustar~dead time. We propagate the
errors in equation \ref{eqn:waveform} using a Monte Carlo simulation.

\begin{figure}
 \includegraphics[height= 8.5 cm,width=6.5cm,trim=0.0cm 0.0cm 0.0cm
0.0cm,clip=false,angle=-90]{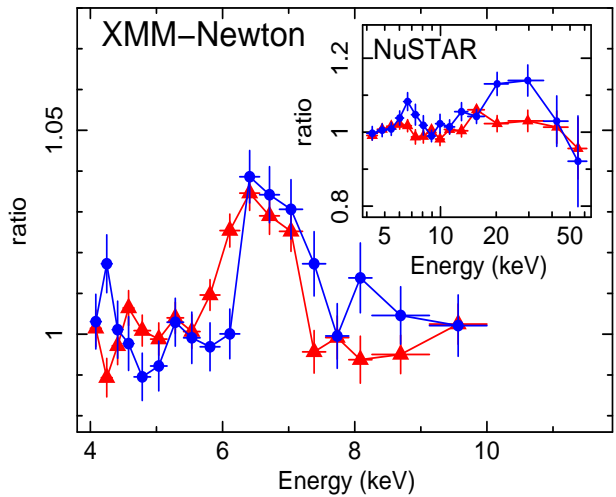}
\vspace{0mm}
 \caption{Spectra from two selected QPO phases, plotted as a ratio to an
absorbed power-law continuum model. The blue circles correspond to a
QPO phase a quarter of a cycle later than the red triangles ($11/16$
cycles compared with $7/16$ cycles). We show \xmm~data averaged
between orbits 2a and 2b. We see shifts in the iron line energy
between the two selected QPO phases, and the hard X-ray coverage of
\nustar~(inset) additionally reveals that the reflection hump is
enhanced relative to the line when the line is blue shifted. Error
bars are 1 $\sigma$.}
 \label{fig:ratio}
\end{figure}

We first reconstruct phase-resolved spectra in the time domain using
equation \ref{eqn:waveform}. We consider 16 QPO phases (i.e. 16 values
of $\gamma$), and throughout we analyse each dataset defined in
Fig. \ref{fig:lc} separately, resulting in five independent
datasets. Fig \ref{fig:ratio} shows examples of the phase-resolved
spectra, plotted as a ratio to an absorbed power-law continuum model
(folded around the telescope response matrix). The continuum model has
been fit ignoring $5.5-8$ keV, where the iron line is prominent and $>
10$ keV, where the reflection hump is prominent. For this plot, we
only consider the \nustar~data and \xmm~orbits 2a and 2b, which were
simultaneous with the \nustar~observation. For plotting purposes, we
have averaged together data from orbits 2a and 2b, even though we
treat them as two separate datasets in our analysis. Red triangles
correspond to a QPO phase of $\gamma=7/16$ cycles and blue circles to
a QPO phase of $\gamma=11/16$ cycles; i.e. the blue points are a
quarter of a cycle after the red points. We see that the line energy
changes over the course of a QPO cycle, and the \nustar~data reveal
that the reflection hump becomes more prominent when the line energy
is higher (blue triangles). In the following section, we model the
iron line with a Gaussian and the continuum with an absorbed power-law
in order to characterise this QPO phase dependence of the line
energy. However, in order to robustly assess the statistical
significance of the line energy modulation, we fit the same model in
the Fourier domain, as described in the following subsection.

\subsection{Phase-resolved spectra in the Fourier domain}

It is straightforward to fit the 16 phase-resolved spectra, as
expressed in equation \ref{eqn:waveform}, with a phenomenological
spectral model to determine if the best-fit spectral parameters vary
systematically with QPO phase. However, assessing the statistical
significance of the spectral parameters is complicated by correlations
between the errors for different QPO phases. For this reason this is
not the method we use to determine significances. Instead, we perform
the fits in the Fourier domain, which provides a different
representation of the same information. The QPO FT, $W_j(E)$, from
equation \ref{eqn:qpoft}, is in units of count rate and, as a complex
quantity, can be expressed in terms of amplitude, $\mu(E)\sigma_j(E)$,
and phase, $\Phi_j(E)$, or in terms of real and imaginary parts,
$\Re\{ W_j(E) \} = \mu(E)\sigma_j(E) \cos[ \Phi_j(E) ]$ and $\Im\{
W_j(E) \} = \mu(E)\sigma_j(E) \sin[ \Phi_j(E) ]$ respectively. The
real and imaginary parts of $W_j(E)$, and the different harmonics, are
statistically independent from one another. Thus standard statistical
methods can be applied if we fit a model to $W_j(E)$ rather than
$w(E,\gamma)$. Here, we first fit spectral models to the
phase-resolved spectra in the time domain to gain insight, before
constructing a model for the QPO Fourier transform. We can exploit the
linearity of the Fourier transform to define a model,
$\tilde{W}_j(E)$, and fold around the telescope response to get the
observed $W_j(E)$, as for a normal spectrum. Specifically, for the
$I^{\rm th}$ energy channel
\begin{equation}
W_j(E_I) = \int_0^\infty \tilde{W}_j(E) R(I,E) dE,
\label{eqn:fold}
\end{equation}
where $R(I,E)$ is the telescope response for the $I^{\rm th}$ energy
channel. We perform a joint fit to real and imaginary parts to
preserve this linearity (which would be lost if we were to instead fit
for amplitude and phase). This results in a joint fit of 5 spectra:
the real and imaginary parts of the first and second harmonics, and
the real part of the DC component (the imaginary part is trivially
zero).

\subsection{Phase difference between harmonics}
\label{sec:psi}

\begin{figure}
 \includegraphics[height= 7.5cm,width=8.5cm,trim=0.8cm 0.0cm 0.0cm
0.0cm,clip=true]{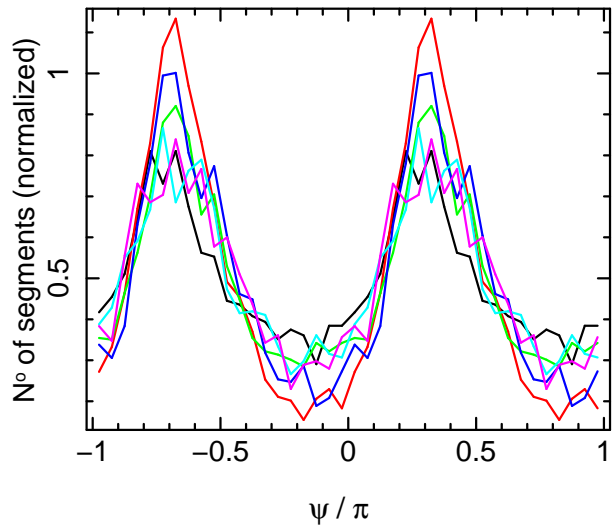}
\vspace{-5mm}
 \caption{Phase difference between the two QPO harmonics, with
   different datasets represented using the same colour scheme as
   Fig. \ref{fig:lc}. For all datasets, we measure the phase
   difference between harmonics $\psi$ for many $32$ s segments (see
   text for details). This plot is a histogram of those measurements
   and shows that there is a well-defined average phase difference
   between the harmonics, which we measure by determining the peak of
   the plotted distribution.}
 \label{fig:dist}
\end{figure}

We first measure the phase difference, $\psi$, between the two QPO
harmonics. This phase difference represents the number of QPO cycles
by which the second harmonic (first overtone) lags the first harmonic
(fundamental), converted to radians (i.e. multiplied by $2\pi$). It is
defined on the interval $0-\pi$ radians, since there are two cycles of
the second harmonic for each cycle of the fundamental. We split the
full band light curve into segments of duration $32$ s. Each segment
contains 512 time bins of duration $dt=0.0625s$, and roughly 8 QPO
cycles. \xmm~orbits 1a, 1b, 2a and 2a contain respectively 2135, 2134,
2455 and 1535 segments with good telemetry, and the
\nustar~observation contains 2217 segments. For each segment, we
calculate the
phase difference $\psi$ following IK15. In Fig. \ref{fig:dist}, we
plot a histogram of these $\psi$ values for each dataset (the colour
scheme is the same as defined in Fig. \ref{fig:lc}), revealing a
strong peak for all datasets. These histograms have two peaks purely
because $\psi$ is cyclical and we show two cycles. We measure the peak
of each histogram following IK15 to obtain the average phase
difference between harmonics. For \xmm~ orbits 1a, 1b, 2a, 2b, we
measure $\psi/\pi=0.309\pm 0.005$, $\psi/\pi=0.336\pm 0.005$,
$\psi/\pi=0.336\pm 0.005$ and $\psi/\pi=0.347\pm 0.006$. For \nustar,
we take the average of the independent measurements for the FPMA and
FPMB to get $\psi/\pi=0.332\pm 0.005$. Note that, even though the
\nustar~observation is simultaneous with orbit 2 of \xmm, the measured
$\psi$ are not required to agree because the full band light curves of
\xmm~and \nustar~cover a different energy range. The agreement we see
between observatories tells us that the phase difference has little
energy dependence here.

\subsection{Energy dependence of QPO amplitude}
\label{sec:rms}

\begin{figure}
 \includegraphics[height= 7.5cm,width=8.5cm,trim=0.8cm 0.0cm 0.0cm
0.0cm,clip=true]{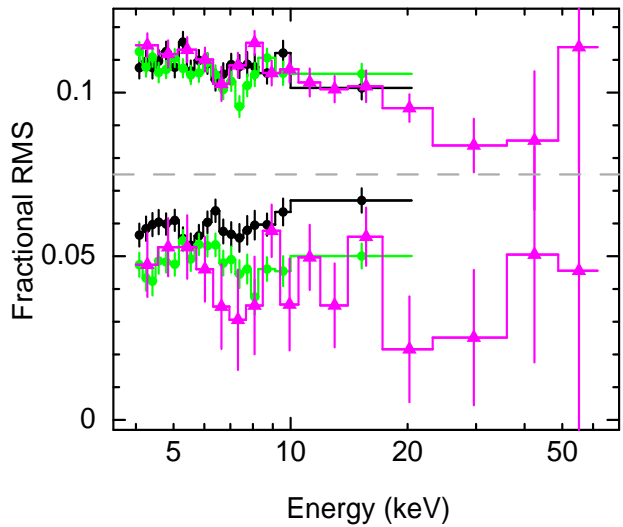}
\vspace{-5mm}
 \caption{
Fractional rms as a function of energy for three selected datasets,
represented using the same colour scheme as Fig. \ref{fig:lc}. The
points above the grey dashed line correspond to the fundamental (first
harmonic) and the points below the dashed line are for the second
harmonic. The other datasets are omitted for clarity. We see tentative
features around the iron line. Error bars are 1 $\sigma$.
}
 \label{fig:rms}
\end{figure}

We measure the fractional rms amplitude of the two QPO harmonics as a
function of energy, $\sigma_j(E)$, for all five datasets. IK15 did
this by calculating the power spectrum for each energy channel. For
the \xmm~data here, we instead calculate the covariance spectrum to
increase signal-to-noise (\citealt{Wilkinson2009}). We follow the
standard procedure for calculating the covariance and its error
(\citealt{Uttley2014}). Our reference band is the full \xmm~band minus
the channel of interest so as to avoid correlating a time series with
itself. For each energy channel, we calculate the cross-spectrum
between that channel and the reference band, and also the power
spectrum of the reference band. The covariance is the modulus squared
of the cross-spectrum divided through by the power spectrum of the
reference band. Since the light curves from each energy channel are
well correlated, the covariance gives a good measure of the power
spectrum with smaller statistical errors
(\citealt{Wilkinson2009}). For \nustar, we circumvent the dead time by
calculating the co-spectrum between the FPMA and FPMB light curve for
each energy channel instead of the power spectrum. Following IK15 we
fit our power spectral estimates (covariance and co-spectrum for \xmm~
and \nustar~respectively) in each energy channel with a sum of
Lorentzian functions. We calculate the fractional rms of each QPO
harmonic from the integral of the corresponding Lorentzian function. A
dead time correction of ${\rm rms}_{\rm det}/{\rm rms}_{\rm in} = 0.8462$ also
must be applied to the \nustar~data. Fig. \ref{fig:rms} shows the
resulting calculation of rms as a function of energy for three of
the five datasets, following the colour scheme of
Fig. \ref{fig:lc}. We show only three datasets to avoid over-crowding
the plot. The points above the dashed line are for the first harmonic,
and below the dashed line are for the second harmonic.

For our Lorentzian fits, we use four Lorentzian functions, one for
each QPO harmonic and two to fit the broad band noise. We tie the
centroid of the second harmonic component to be double that of the
first harmonic and force the two QPO Lorentzians to have the same
quality factor ($Q =$ centroid frequency / full width at half
maximum). The centroid and quality factor of the QPO fundamental
component are free to vary with energy, but we measure no significant
energy dependence for either of these quantities. We tried many
variations on the model to test for the robustness of the fit. We
tried using more and less broad band noise Lorentzians, allowing the
QPO components to have different quality factors, relaxing the
centroid frequency ratio of 2, fixing the widths and/or centroid
frequencies of the QPO Lorentzians to equal those measured for the
full band and so on. We even tried simply integrating the power
spectral estimates over the widths of the QPO components instead of
fitting a model. In all cases, we obtained consistent results,
indicating that our fits are robust.

\subsection{Phase lags between energy bands}
\label{sec:lags}

\begin{figure}
 \includegraphics[height= 9.5cm,width=8.5cm,trim=0.8cm 0.0cm 0.0cm
0.0cm,clip=true]{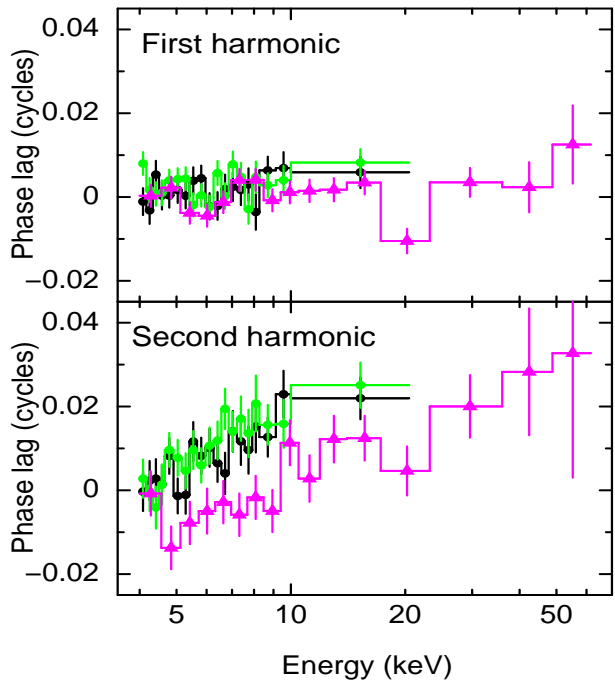}
\vspace{-5mm}
 \caption{Phase lag of each energy channel relative to a reference
   band for three selected datasets, represented using the same colour
   scheme as Fig. \ref{fig:lc}. The reference band is the full band of
   the respective instrument, and therefore is slightly different
   between \xmm~and \nustar. This creates the small offset seen in
   the second harmonic. As with Fig. \ref{fig:rms}, the other datasets
   are omitted for clarity. Error bars are 1 $\sigma$.
}
 \label{fig:lags}
\end{figure}

We calculate the phase lag between each energy channel and a broad
reference band for both QPO harmonics, $\Delta_j(E)$. For \xmm, we use
the same reference band as described above for the covariance
spectrum. We calculate the cross-spectrum for each channel and average
this over the width of each QPO harmonic, as defined by the Lorentzian
fitting described in the previous section. The phase lag for each QPO
harmonic is the argument of this averaged cross-spectrum. For \nustar,
we again utilise the two independent focal plane modules. We use the
full FPMB band as the reference band and calculate the cross-spectrum
between this and each channel of interest in FPMA. We also calculate
an independent set of cross-spectra using FPMA as the reference band
and FPMB for the subject bands. For each energy channel, we average
together these two independent measurements of the cross-spectrum to
increase signal-to-noise. Fig. \ref{fig:lags} shows the lag spectra
for the same three datasets as the previous plot. We see a small
offset in the second harmonic between \xmm~and \nustar. This is simply
because the lag spectra are calculated for \nustar~using a different
reference band, and the lag of the second harmonic depends on
energy. This will introduce a small offset between \xmm~and
\nustar~when it comes to plotting best-fit spectral parameters against
QPO phase. As it turns out, this offset is small enough to ignore
completely, but even if it were large, it would be fairly simple to
correct for since it is just a constant offset. We calculate the phase
offsets $\Phi_1(E)$ and $\Phi_2(E)$ using equation \ref{eqn:Phi}.

\subsection{Step-by-step summary}

The steps of the IK15 method can be summarised as follows:
\begin{enumerate}
\item Measure the phase difference between the QPO harmonics in a
  broad reference band (see Section \ref{sec:psi}),
\item For both QPO harmonics, measure the rms variability as a function
  of energy (see Section \ref{sec:rms}),
\item For both QPO harmonics, measure the phase-lag between each energy
  channel and the reference band (see Section \ref{sec:lags}),
\item Combine these measurements in order to calculate the QPO FT
  (Equation \ref{eqn:qpoft}),
\item Inverse FT to obtain a waveform for each energy channel
  (Equation \ref{eqn:waveform}.
\end{enumerate}
For the Fourier domain method, we stop at step 4 and fit a model
directly to the QPO FT, whereas the time domain method also includes
step 5.

\section{Results}
\label{sec:results}

Fig. \ref{fig:ratio} shows the spectrum for two selected QPO phases
plotted as a ratio to an absorbed power-law continuum model, with the
blue circles representing the spectrum a quarter of a cycle later than
the red triangles. We see a shift in line energy between the two QPO
phases. In this section, we fit the iron line with a Gaussian function
to characterise the phase dependence of the centroid energy and assess
its statistical significance.

\subsection{Time domain fits}

We first fit the phase-resolved spectra in the time domain with an
absorbed power-law plus Gaussian model, in the energy range
4-10 keV. We consider the five datasets separately, which allows us to
compare results for independent analyses. We initially tie
all spectral parameters to remain constant during the QPO cycle and
test if the fit is improved when we allow each parameter to vary
freely with QPO phase. For all three datasets, we achieve the minimum
reduced $\chi^2$ value by allowing the Gaussian centroid energy
($E_{line}$), Gaussian flux ($N_G$) and the power-law index
($\Gamma$) and normalisation ($N_{cont}$) to vary with QPO phase. The
fit is not improved by allowing the Gaussian width to vary with QPO
phase. We plot the best-fit line centroid energy against QPO phase
(light blue circles) in Figure \ref{fig:sepmod}. We do not plot error
bars here, since the errors are correlated between QPO phases in the
time domain fits. All datasets show a modulation in the line
energy. For all but orbit 1b of XMM-Newton, the line energy modulation
has the same distinctive shape, with maxima at $\sim 0.2$ and $\sim
0.7$ cycles. The modulations in $\Gamma$, $N_G$ and $N_{cont}$ (not
pictured) are also consistent between these 4 datasets.

It is puzzling that orbit 1b disagrees with the other datasets. This
dataset also exhibits different modulations in $N_G$ and $\Gamma$ from
the others (this can be seen in Fig. \ref{fig:allmod}, which is
explained in detail in the following sections). To investigate this further, we
split up orbit 1 into four quarters such that the first two quarters
together make up orbit 1a and the final two quarters together make up
orbit 1b. We find, as expected, that the first two quarters both show
the same modulation in line shape seen for orbit 1a. The fourth
quarter (i.e. the second half of orbit 1b) shows a peak in line energy
at $~0.7$ cycles but not at $0.2$ cycles, so is different from orbit
1a but only slightly. This fourth quarter also shows a modulation in
$\Gamma$ consistent with orbit 1a. It is the third quarter that
differs so radically from all of the other datasets. This shows a peak
in line energy at $\sim 0.4$ cycles, and also exhibits a different
(but very weak) modulation in $\Gamma$ from orbit 1a. The fact that
the two halves of orbit 1a display the same modulations as each other,
and as orbit 1a treat as a whole, gives us confidence in the
robustness of the method, and implies that something different
really is happening in this third quarter of orbit 1.

\subsection{Fourier domain fits}

We now fit the same phenomenological spectral model directly to the
QPO FT derived from the data. This will allow us to
assess the statistical significance of the line energy modulation. We
construct a model for the QPO Fourier transform by representing the
spectral parameters as periodic functions of QPO phase, $\gamma$. For
example, the line energy is
\begin{equation}
E_{line}(\gamma) = E_0 + A_{1E} \sin[ \gamma - \phi_{1E} ] + A_{2E}
\sin[ 2(\gamma - \phi_{2E}) ],
\label{eqn:eline}
\end{equation}
where $E_0$, $A_{1E}$, $A_{2E}$, $\phi_{1E}$ and $\phi_{2E}$ are model
parameters. We see that $E_0$ is the mean line energy, and all
variability in the line energy as a function of QPO phase is captured
by the amplitudes and phases of the sine waves. The other potentially
varying spectral parameters (Gaussian width and normalisation,
power-law index and normalisation) are also modelled in the same
manner with 5 parameters each. Our model calculates the resulting
spectrum for 16 QPO phases and then calculates the Fourier transform
for a grid of energy bins. We then fold, for each harmonic, the real
and imaginary parts of this Fourier transform around the telescope
response matrix (equation \ref{eqn:fold}) and fit to the observed QPO
Fourier transform (equation \ref{eqn:qpoft}).

\subsubsection{Separate fits}

\begin{figure}
\begin{center}
 \includegraphics[height= 9.9cm,width=8.0cm,trim=0.0cm 0.0cm 0.0cm
0.0cm,clip=false]{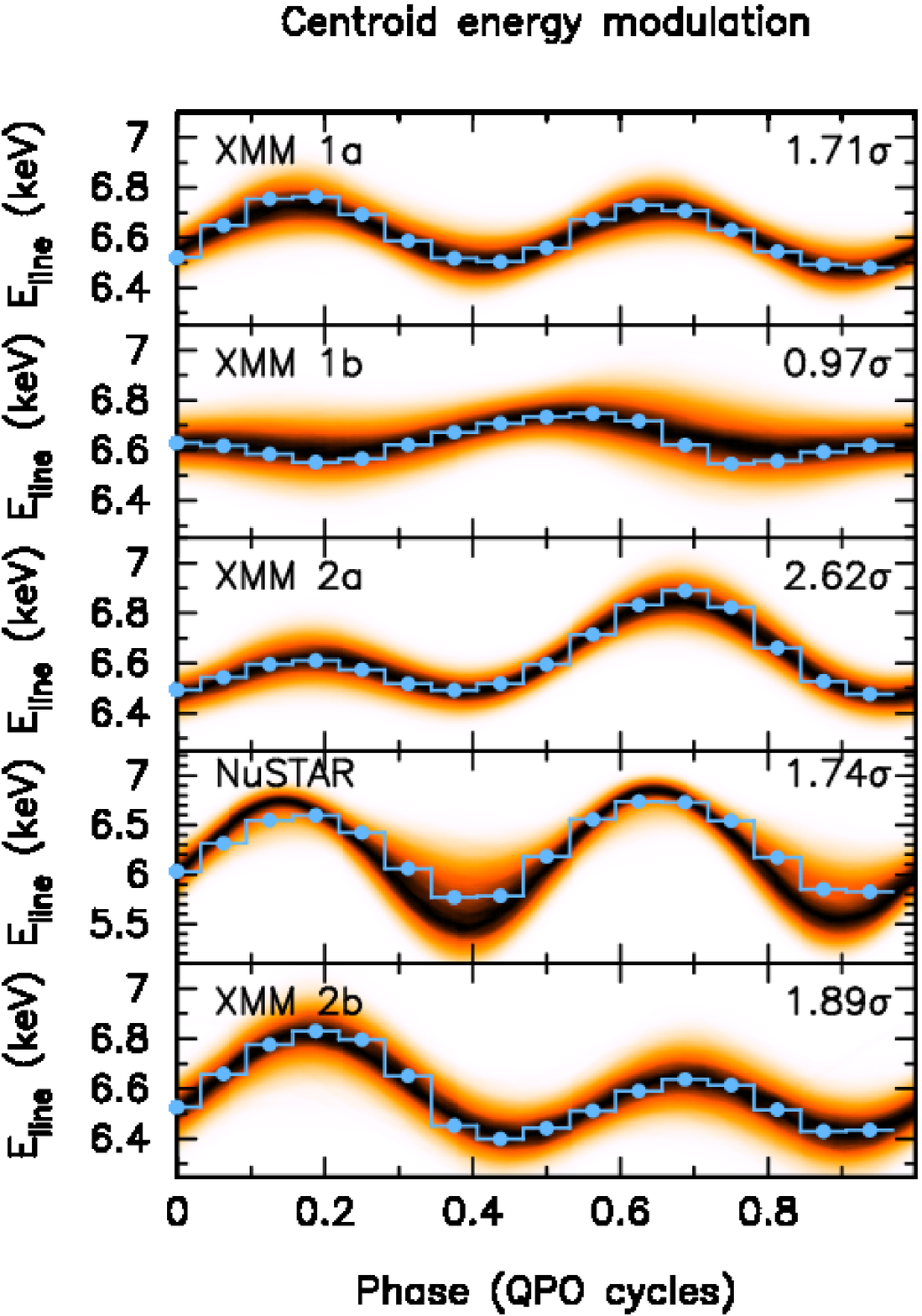} \\
~~~~~~~~~~~~~~~~~~~~\includegraphics[height=6.0cm,width=1.5cm,trim=0.0cm 0.0cm 0.0cm
 0.0cm,clip=false, angle=-90]{legend.ps}
\end{center}
\vspace{0mm}
 \caption{
Iron line centroid energy as a function of QPO phase for each of the
five datasets (as labelled). Light blue circles are the results of our
time domain fits and the probability maps are the results of our
Fourier domain fits. For the time domain fits, we fit an absorbed
power-law plus Gaussian model to spectra corresponding to 16 QPO
phases. For the Fourier domain fits, we consider the same model, with
parameters varying periodically with QPO phase, and FT the model to
fit to the data in Fourier space. We determine the significance of the
modulations (as labelled) from the Fourier domain fits, and create the
probability maps using a Monte Carlo Markov Chain (see text for
details). The maps are normalised such that they peak at unity, and
the colours are defined in the key. Note that $P/P_{\rm max}=0.1$,
below which the colour scale looks rather white, corresponds to the
$\sim 2.15 \sigma$ confidence contour.}
 \label{fig:sepmod}
\end{figure}

As with the time domain fits, we fit the five datasets separately,
expecting to see exactly the same results as before (since the FT of a
function is simply a different representation of the same function),
but with more manageable statistics. We again find a best-fit with
modulations in the line energy (i.e. $A_{1E}>0$ and $A_{2E}>0$) and
flux and the continuum normalisation and power-law index, and again
the fits are not improved by allowing the Gaussian width to vary with
QPO phase. As an example of our fits, we plot in Fig. \ref{fig:qpoft} (left)
the QPO FT for orbit 1a (black points) along with the best-fit model
(lines). Here, the data are unfolded around the instrument response
assuming the best-fit model and are in units of energy squared $\times$
specific photon flux (i.e. the \textit{eeuf} option in
\textsc{xspec}). The best-fit model for the first harmonic is plotted
in red and the second harmonic in blue. We see features in the data
and model around the iron line, which result in the model from
modulations of the line energy and equivalent width. The second
harmonic shows the clearest features, with an excess at $\sim 6.2$ keV
in the real part and a dip at the same energy in the imaginary part,
surrounded by two peaks either side. We fit jointly for the real and imaginary parts of
both harmonics, and also for the mean spectrum (the real part of the
DC component) which is not pictured here.

We plot the $E_{line}(\gamma)$ function derived from our Fourier
domain fits, visualised as a probability map, in
Fig. \ref{fig:sepmod}. The best-fit function $E_{line}(\gamma)$ can be
plotted by substituting the best-fit values for $E_0$, $A_{1E}$,
$A_{2E}$, $\phi_{1E}$ and $\phi_{2E}$ into equation
\ref{eqn:eline}. Here, we also take into account the probability
distributions of these 5 parameters by running a Monte Carlo Markov
Chain (MCMC) in \textsc{xspec} and then, for each step in the chain,
calculating the $E_{line}(\gamma)$ function. We then create a
histogram to plot the posterior distribution for the function. Details
of the chain and of the calculation of these histograms are presented
in Appendix \ref{sec:datavis}. As expected, the frequency domain results plotted in
Fig. \ref{fig:sepmod} agree with the time domain fits (light blue
circles), but we are now able to visualise the uncertainty on the
best-fit line energy modulation (see the key).

We calculate the significances quoted in Fig. \ref{fig:sepmod} by
comparing the $\chi^2$ from the best fit model for each dataset with
the minimum $\chi^2$ achieved for the same dataset when the line
energy amplitudes are fixed to $A_{1E}=A_{2E}=0$. This null hypothesis
model has 4 more degrees of freedom than the best fit model, because
it is insensitive to the phase parameters, $\phi_{1E}$ and
$\phi_{2E}$. We compare the best-fit to the null hypothesis using an
f-test, converting p-values to sigmas in the standard way
(e.g. $1\sigma$ corresponds to $p=0.317$).

\subsubsection{The case of orbit 1b}

\begin{figure*}
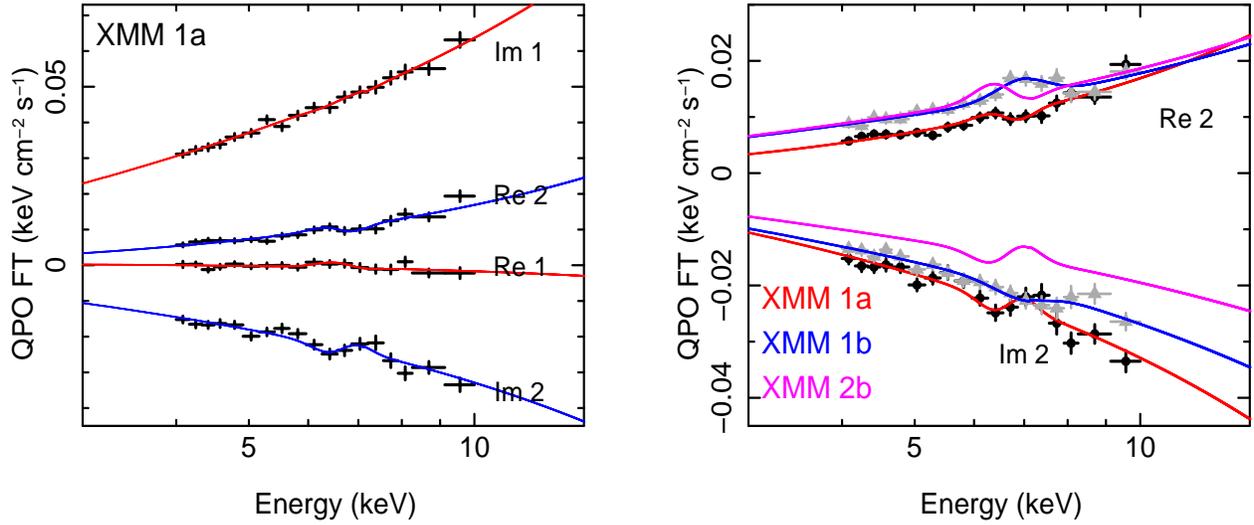

 \includegraphics[height= 7.5cm,width=8.5cm,trim=0.0cm 0.0cm 0.0cm
0.0cm,clip=true]{QPOFTplot.ps} ~~~
 \includegraphics[height= 7.5cm,width=8.5cm,trim=0.0cm 0.0cm 0.0cm
0.0cm,clip=true]{FTcompare.ps}
\vspace{-5mm}
 \caption{\textit{Left:} QPO Fourier Transform as a function of energy
   for \xmm~orbit 1a. Real and imaginary parts of the first and
   second harmonics are as labelled. Here, we plot the data (black
   points) unfolded around the instrument response matrix, assuming
   the best-fit model (red lines for the first harmonic and blue lines
   for second harmonic), in units of energy squared $\times$ specific photon
   flux (i.e. the \textit{eeuf} option in
   \textsc{xspec}). \textit{Right:} QPO FT for the anomalous
   \xmm~orbit 1b (grey triangles and blue lines) compared with
   \xmm~orbit 1a (black circles and red lines), only considering the second
   harmonic (real and imaginary parts as labelled). We see clear
   differences in the shape for both real and imaginary parts. To
   demonstrate that \xmm~orbit 1a is representative of all the other
   datasets, we also plot the best-fit model for orbit 2b. Error bars
   are 1 $\sigma$.}
 \label{fig:qpoft}
\end{figure*}

As with the time domain fits, it is striking that all datasets except
for orbit 1b show the same characteristic trend, with peaks in line
energy at $\sim 0.2$ and $\sim 0.7$ cycles. Looking at the QPO FT
reveals that the difference between orbit 1b and all the other
datasets is in the second harmonic. Fig. \ref{fig:qpoft} (right) shows
the FT of the second harmonic (real and imaginary parts as labelled)
for orbit 1a (black circles) and orbit 1b (grey triangles). The best
fit models for orbits 1a and 1b are plotted in red and blue
respectively. We see very different behaviour between the two
datasets. Where orbit 1a shows a dip (real part at $\sim 7$ keV),
orbit 1b shows an excess. Where orbit 1a shows an excess (imaginary
part at $\sim 7$ keV), orbit 1b shows a dip. All other datasets
display similar behaviour to orbit 1a. To illustrate this, we plot the
best-fit model for orbit 2b in magenta. This has a slightly different
normalisation, but the same characteristic shape as orbit 1a.

We check if these differences can result from our assumptions when
measuring the fractional rms as a function of energy. For the many
different methods of measuring this described in section
\ref{sec:rms}, we measure QPO FTs consistent with before and therefore
obtain results consistent with Fig. \ref{fig:sepmod}. We therefore
conclude that the method produces robust results and that orbit 1b
really does seem to be doing something different to the other datasets.

\subsubsection{Joint fits}

Since all datasets show a modulation in line energy, we combine them
into a joint fit to compare with the null hypothesis:
$A_{1E}=A_{2E}=0$. We first leave out the anomalous dataset, orbit
1b. We see in Fig. \ref{fig:sepmod} that the two maxima in line energy
measured for \nustar~slightly lead those measured for the
simultaneous \xmm~orbit 2. This is because we used a different
reference band for \nustar. Since this constant offset turns out to be
very small in this case, we are able to ignore it. We therefore tie
the modulations in line energy, line flux and power-law index to be
the same for all four considered datasets, but allow the power-law
normalisation to differ for different datasets. We note that the
modulation in power-law normalisation is very similar for each dataset
(even including orbit 1b), but is well constrained enough for small
differences in datasets to be highly statistically significant. We tie
the power-law index between \xmm~and \nustar~using the formula
$\Gamma_{NuSTAR}(\gamma) = \Gamma_{XMM}(\gamma) + \Delta \Gamma$, in
order to account for the cross-calibration discrepancy. Similarly, we
tie the line energy between observatories using the formula
$E_{line,NuSTAR}(\gamma) = C E_{line,XMM}(\gamma)$. We use
$\Delta\Gamma=0.236$ and $C=0.970$. We obtain a good
fit (reduced $\chi^2 = 287.13 /279 = 1.029$) with the differential
properties of the spectral parameters consistent with before.

When we also include the orbit 1b data in our fit, tying
all parameter modulations except for the power-law normalisation
across all datasets, we obtain a fit with reduced
$\chi^2=370.32/364=1.017$. When we allow $\phi_{1E}$ and $\phi_{2E}$
to be different for orbit 1b compared with all the other datasets (as 
seems to be the case from Fig. \ref{fig:sepmod}), the fit improves with
reduced $\chi^2 = 361.82/362=1.000$. An f-test determines that this is
a $2.43\sigma$ improvement, indicating that the line
energy modulation in orbit 1b is likely different from the other
datasets. Also freeing $A_{1E}$ and $A_{2E}$ for orbit 1b does not
further improve the fit, so we keep these amplitudes tied across all
datasets. When we also allow the second harmonic amplitudes of the
$N_G$ and $\Gamma$ modulations to be different for orbit 1b, the fit
again improves, with reduced $\chi^2=354.24/360=0.984$. An f-test
indicates that this is a $2.29\sigma$ improvement. This is our
best-fit model.

\begin{figure*}
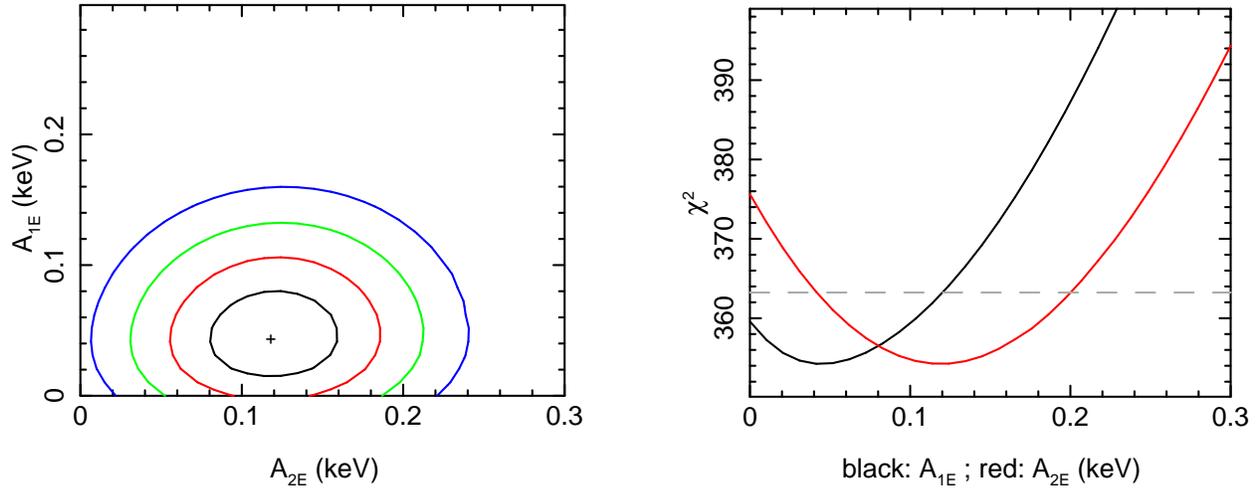

 \includegraphics[width=8.5cm]{ref_everything_A1A2contours.ps} ~~~
 \includegraphics[width=8.5cm]{ref_everything_seperatecontours.ps}
\vspace{-5mm}
 \caption{$\chi^2$ contour plots from our Fourier domain fits
   considering all datasets. We show the amplitude of the first and
   second harmonic of the line energy modulation $A_{1E}$ and
   $A_{2E}$. \textit{Left:} Two dimensional contour plot. The cross
   denotes the best-fit and the black, red, green and blue lines
   correspond to 1, 2, 3 and 4 $\sigma$ confidence contours
   respectively for two degrees of freedom. \textit{Right:} One
   dimensional plot for $A_{1E}$ (black) and $A_{2E}$ (red). The grey
   dashed line is $\Delta\chi^2=9$ ($3\sigma$ for one degree of
   freedom).}
 \label{fig:contours}
\end{figure*}

\begin{table}
	\centering
	\caption{Best fit line energy parameters for our joint
          fit. Errors are $1\sigma$.}
	\label{tab:paras}
  \bgroup
  \def\arraystretch{1.5}
	\begin{tabular}{lcc} 
		\hline
		  Parameter & \vline & Best fit \\
		\hline
                $A_{1E}$ (keV) & \vline & $0.0446^{+0.023}_{-0.020}$  \\
              $\phi_{1E}$ (cycles) & \vline & $0.373^{+0.076}_{-0.13}$ \\
                $A_{2E}$ (keV) & \vline & $0.119^{+0.026}_{-0.026}$  \\
              $\phi_{2E}$ (cycles) & \vline & $0.0497^{+0.019}_{-0.018}$ \\
                $E_0$ (keV) & \vline & $6.60^{+0.019}_{-0.018}$  \\
		\hline
	\end{tabular}
  \egroup
\end{table}

Our best-fit parameters for the line energy modulation are presented
in Table \ref{tab:paras}. Fig. \ref{fig:contours} (left) is a contour plot
resulting from varying $A_{1E}$ and $A_{2E}$ (using the
\textit{steppar} command in \textsc{xspec}). The contours represent
$\Delta \chi^2=$ $2.3$ (black), $6.18$ (red), $11.83$ (green) and
$19.33$ (blue). These $\chi^2$ levels correspond to 1, 2, 3 and 4
$\sigma$ confidence for two degrees of freedom. We see that a fairly
large part of parameter space can be ruled out with $4\sigma$
confidence. The null hypothesis model ($A_{1E}=A_{2E}=0$), now has 6
more degrees of freedom than the best-fit model, because the null
hypothesis model is insensitive to $\phi_{1E}$ and $\phi_{2E}$ for
orbit 1b, plus the same two parameters for the other datasets. We
compare the best-fit achieved by fixing $A_{1E}=A_{2E}=0$
($\chi^2=380.68/366$) with our global best-fit ($\chi^2=354.24/360$)
using an f-test, which rules out the null-hypothesis with $3.70\sigma$
confidence.

Fig \ref{fig:contours} (right) shows $\chi^2$ plotted against $A_{1E}$
(black) and $A_{2E}$ (red). The dashed line depicts $\Delta \chi^2 =
9$, which corresponds to $3\sigma$ for 1 degree of freedom. We see
that $A_{2E}$ in particular is fairly well constrained, with $3\sigma$
confidence limits of approximately $0.04 < A_{2E} < 0.21$. The
best-fit achieved when fixing $A_{1E}=0$ has reduced
$\chi^2=359.58/363$ and the best-fit achieved with $A_{2E}=0$ is 
$\chi^2=375.63/363$. Comparing these to the global best fit yields
significances of $1.46\sigma$ for the first harmonic and $3.89\sigma$
for the second harmonic.

\begin{figure*}
\begin{center}
 \includegraphics[height= 9.9cm,width=8.0cm,trim=0.0cm 0.0cm 0.0cm
0.0cm,clip=false]{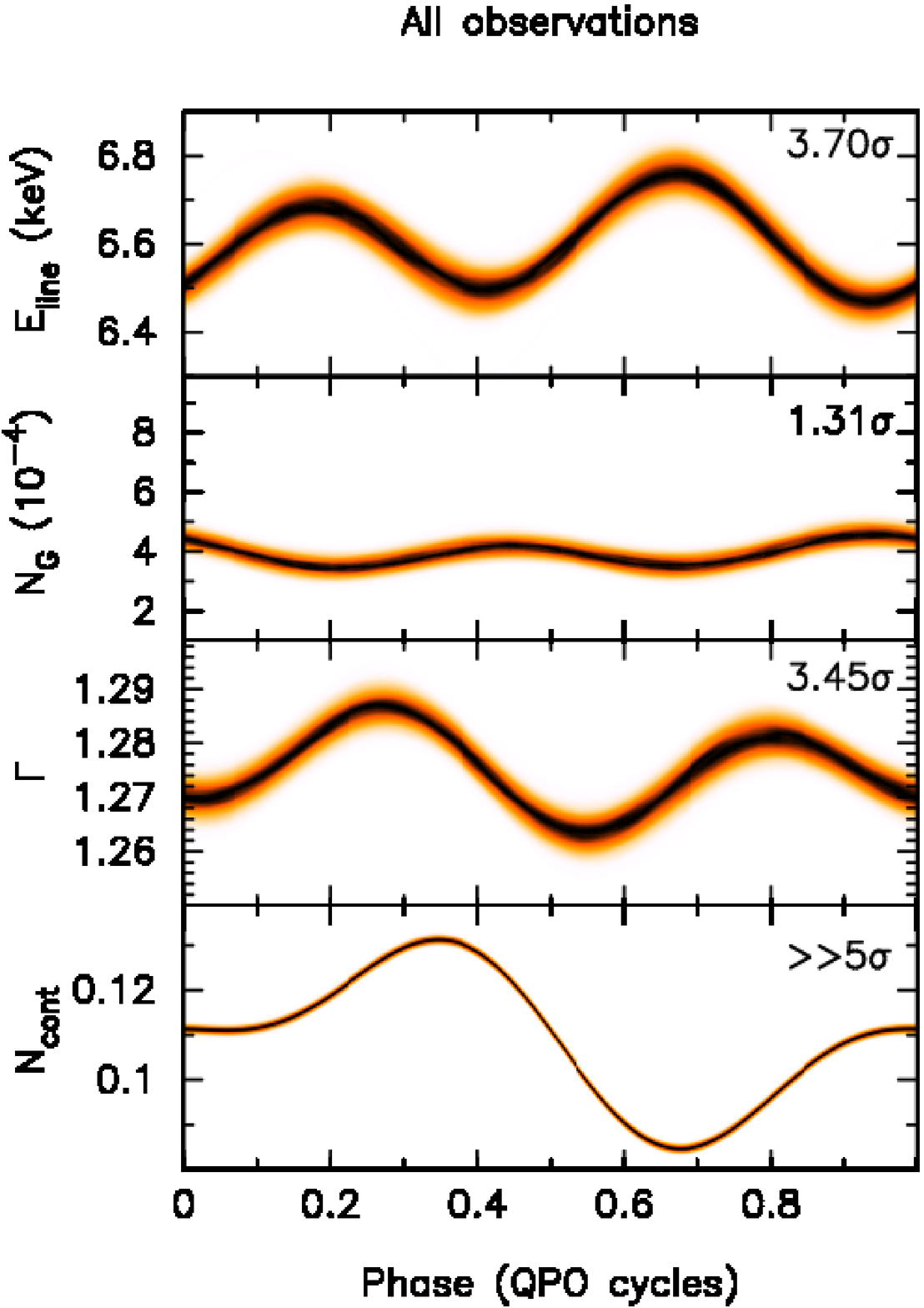} ~~~
 \includegraphics[height= 9.9cm,width=8.0cm,trim=0.0cm 0.0cm 0.0cm
0.0cm,clip=false]{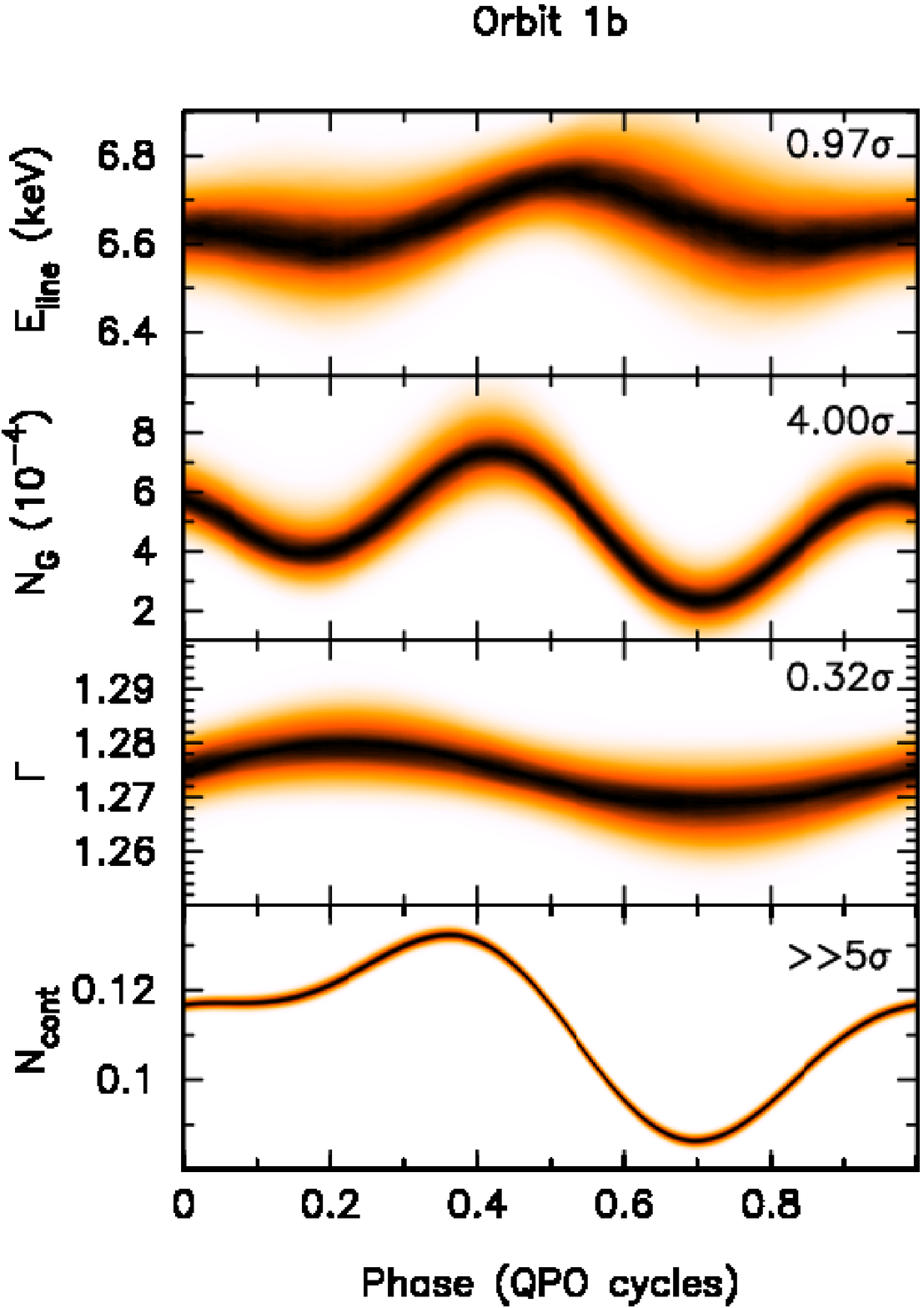}
\end{center}
\vspace{0mm}
 \caption{Probability maps from our Fourier domain fits for line
   energy ($E_{line}$), line flux ($N_G$), power-law index ($\Gamma$)
   and power-law normalisation ($N_{cont}$). Significances for each
   parameter are as labelled and colours are as defined in
   Fig. \ref{fig:sepmod}. \textit{Left:} The results of our joint
   fits, considering all datasets. For parameters which are not tied
   across all datasets (see text for details), we plot the values
   corresponding to orbit 1a. \textit{Right:} The results for only the
   anomalous orbit 1b. We see clear differences from the other
   datasets, with perhaps the most striking being the large modulation
   in iron line flux.}
 \label{fig:allmod}
\end{figure*}

Fig. \ref{fig:allmod} (left) shows the probability map for all four
variable spectral parameters for our final joint fit. Here, for
parameters which are not tied across all datasets (such as
$\phi_{1E}$ and $\phi_{2E}$), we plot the values corresponding to
orbit 1a. Note that the functions $E_{line}(\gamma)$, $N_G(\gamma)$
and $\Gamma(\gamma)$ are tied across \textit{all} datasets except for
orbit 1b. We see no statistically significant modulation in the iron
line flux, but we do see a modulation in the power-law index which
lags the line energy modulation by $\sim 0.1$ cycles. In
Fig. \ref{fig:allmod} (right), we make the same plot for the case of
orbit 1b. Even though the statistics are of course worse, the parameter
modulations are strikingly different. The line energy and power-law
index are both consistent with being constant, but the iron line flux
varies with a large amplitude and high statistical
significance. Clearly, there is something very different about orbit
1b. We have checked for proton flares, absorption events and various
instrumental issues, but find no contribution from these effects, so
are forced to conclude that this anomalous behaviour during orbit 1b
is intrinsic to the source. We note that the iron line width is larger
during orbit 1b ($0.51 \pm 0.05$keV) than for the other datasets
combined ($0.43 \pm 0.02$ keV). We can also see in the $4-10$ keV
power spectrum (Fig. \ref{fig:psd}) a slight increase in the amplitude
of the fundamental from orbit 1a (black) to 1b (red), but a very
slight decrease in the amplitude of the second harmonic. Also,
  the broad band noise above $\sim 2$ Hz changes a little between
  orbit 1a and 1b. These differences may be indicative of their being
a slightly different geometry during orbit 1b.

\section{Discussion}
\label{sec:discussion}

\begin{figure}
\begin{center}
 \includegraphics[height= 10.5cm,width=7.5cm,trim=0.0cm 1.0cm 20.0cm
0.0cm,clip=false]{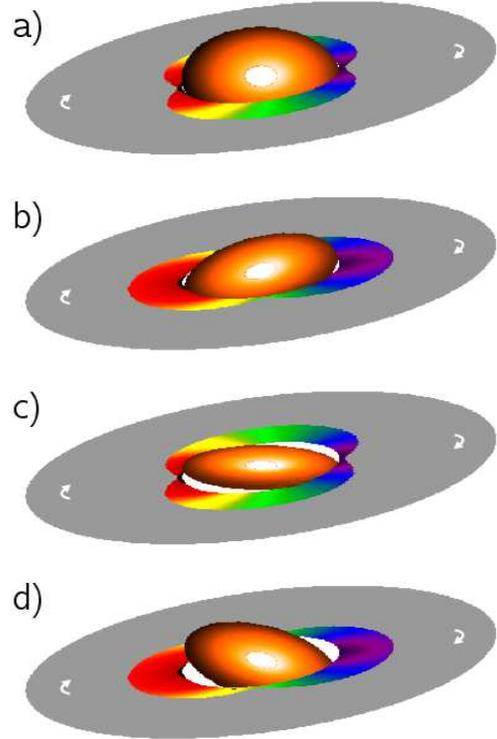}
\end{center}
\vspace{0mm}
 \caption{Schematic representation of the precessing inner flow
   model. The inner flow (orange) extends out to $\sim 20-30~R_g$ and
   is misaligned with both the disk (grey) and black hole equatorial
   plane (horizontal). The flow precesses around the (vertical) black
   hole spin axis such that the front of the flow faces us in (a), to
   our left in (b) and so on. The front and back of the flow irradiate
   the disk, illustrated here by the multi-coloured patches. As the
   flow precesses, these irradiated patches rotate over the disk
   surface, pro-grade with disk orbital motion (white arrows). The
   colours of the irradiated patches encode energy shifts due to disk
   orbital motion and gravitational redshift.}
 \label{fig:schem}
\end{figure}

We have further developed the QPO phase-resolving method of IK15 and
applied it to, in total, $\sim 260$ ks of \xmm~data and $\sim 70$ ks
of \nustar~data from the 2014 outburst of H 1743-322. We measure a
statistically significant ($3.7 \sigma$) modulation of the iron line
centroid energy with QPO phase by combining five independent
datasets. We see in Fig. \ref{fig:sepmod} that, for four of the five
datasets, the line energy modulation has the same distinctive shape,
with maxima at $\sim 0.2$ and $\sim 0.7$ QPO cycles. Surprisingly, one
dataset (\xmm~orbit 1b) does not show the same trend. Here we discuss
the implications of the measured modulation and speculate as to why
orbit 1b differs from the other datasets.

\subsection{Interpretation: Precession}

Our result provides strong evidence that the Type C QPO observed here
is driven by systematic changes in the accretion geometry over the
course of a QPO cycle. The only mechanism by which the line energy can
vary without a geometric change is through shifts in the rest frame
line energy driven by changes in the disk ionisation state. An
increase in disk ionisation increases the iron line energy and
\textit{suppresses} the flux in the reflection hump relative to the
line (e.g. \citealt{Ross2005}; \citealt{Garcia2013}), in conflict with
what we observe (Fig. \ref{fig:ratio}). Also, the observed $\sim 6.4$
keV to $\sim 6.8$ keV change in line energy would require a factor
$\sim 200$ change in illuminating flux over a QPO cycle to originate
purely from variations of disk ionisation (see e.g. Fig. 1 in
\citealt*{Matt1993}); which is implausible for all datasets except for
orbit 1b, which show a change in line flux smaller than a factor of 2
(see Fig. \ref{fig:allmod}, left). This indicates that the line energy
variation is driven, at least in part, by changes in the relativistic
distortions to the iron line profile, and therefore by a geometric
variation over a QPO cycle. This ties in with recent population
studies (\citealt{Motta2015}; \citealt*{Heil2015}) which show that
systems observed with a more edge-on disk display systematically
higher amplitude Type C QPOs.

Shifts in the line energy are predicted to arise if the QPO originates
from Lense-Thirring precession of the hot inner flow
(\citealt*{Ingram2009}; \citealt{Ingram2012a}). As the inner flow
precesses, it preferentially illuminates different disk azimuths,
giving rise to a blue/red shifted iron line when the
approaching/receding disk material is irradiated. For a geometry in which a single
bright patch rotates about the disk surface, completing one cycle per
QPO cycle, we would observe one maximum and one minimum in line energy
per QPO cycle. Instead, we observe two maxima, neither of which
coincide with a peak in continuum flux. This can be explained if we
consider two bright patches rotating about the disk surface, as
illustrated in Fig. \ref{fig:schem}. In this picture, the inner flow (orange)
precesses, but the disk (grey) is held stationary by viscosity
(\citealt{Bardeen1975}). The disk transitions into the hot inner flow
at the truncation radius. In this schematic, the disk is irradiated by
both the front and back of the flow (see the multi-coloured
patches), as we may expect to happen if the inner flow is sufficiently
thin for its underside to be above the disk mid-plane (or for a very
large misalignment between the disk and inner flow). The calculations
of \cite{Ingram2012a} considered an inner flow with very large
vertical extent, and therefore only predicted one bright patch on the
disk, as the underside of the flow was never above the disk
mid-plane. The Doppler shifts experienced by photons reflected from
respectively approaching and receding disk material are illustrated in
Fig. \ref{fig:schem} by the colour scheme of the irradiated patches. Precession of
the flow as illustrated in Fig. \ref{fig:schem} predicts a rocking of the iron
line shape twice per precession cycle as different disk azimuths are
illuminated first by the front of the flow, then half a cycle later by
the back. The maximum line energy will occur when the approaching and
receding sides of the disk are illuminated (Fig. \ref{fig:schem} b and c), since
Doppler boosting means that the blue shifted part of the line (the
so-called `blue horn') will dominate over the red shifted part (the
so-called `red wing').

Non-relativistic precession mechanisms are unlikely. Classical
precession is expected around an oblate spinning star but not for a
black hole (\citealt{Stella1998}). Magnetic precession can result when
the magnetic field of a spinning star intersects the accretion flow
(\citealt{Shirakawa2002}), but not for astrophysical black holes
which, without electric charge, have no way to generate their own
magnetic field. Radiation pressure can cause variable warping in the
outer disk through non-linear growth of perturbations, but only at
disk radii $\gtrsim 160~R_g$ (\citealt{Pringle1996};
\citealt*{Frank2002}), where orbital motion is too slow to explain the
large observed energy shifts in the line. It is therefore likely that
we are specifically witnessing Lense-Thirring precession. We note that
Lense-Thirring precession of the reflector (the disk) rather than the
illuminator (the inner flow) could potentially reproduce the observed
line energy modulation (\citealt*{Schnittman2006};
\citealt{Tsang2013}); although we note that the QPO modulates the
power-law spectrum emergent from the inner flow much more strongly
than the thermal disk emission visible at low energies. We also note
that the observed line energy modulation could potentially result from
a precessing jet (\citealt{Kalamkar2016}).

For Lense-Thirring
precession of the entire inner flow, the precession period depends on
the inner and outer radii of the inner flow, the radial surface
density profile of the inner flow (\citealt{Fragile2007};
\citealt*{Ingram2009}), as well as the mass and dimensionless spin
parameter, $a=cJ/GM^2$, of the black hole. Assuming a constant surface
density, a canonical black hole mass of $10~M_{\odot}$ and a spin of
$a=0.2$ (\citealt{Steiner2012}; \citealt{Ingram2014}), the $\sim 4$ s
period implies a truncation radius of $\sim 20-30~R_g$.

\subsection{Implications}

Lense-Thirring precession arises (due to the General Relativistic
frame dragging effect) only in orbits with their rotational axis
misaligned with the black hole spin axis. This may occur for accreting
material in binary systems in which the black hole spin axis is
misaligned with the axis of binary orbital motion (as the result of an
asymmetric natal supernova kick; \citealt{Fragos2010}). Quite how the
accretion flow reacts to this misalignment is a challenging
theoretical question, which will be informed by our result. For a
classical thin disk, the inner regions
have long been thought to align with the black hole and the outer
regions with the binary (\citealt{Bardeen1975}), but the location of
the transition between orientations has remained uncertain. Recent
simulations (\citealt{Krolik2015}) find this radius to be $\sim
8-9~R_g$, which is small enough to be within the disk truncation
radius of $\sim 20-30~R_g$ indicated here by setting the precession period
equal to the QPO period. This implies that the inner flow is being fed
by material from the truncated disk out of the black hole equatorial
plane (as in Fig. \ref{fig:schem}). Grid-based General Relativistic
magnetohydrodynamic simulations of accretion flows in
which the vertical extent is large compared with viscosity indicate
that the entire hot inner flow can precess in this situation due to
strong coupling through pressure waves (\citealt{Fragile2007}), in
line with what is illustrated in Fig. \ref{fig:schem}. Alternatively, calculations
using an $\alpha$-prescription viscosity with a large misalignment
angle and/or low viscosity show evidence for the disk breaking into
discrete, independently precessing rings (\citealt{Nixon2012}). This
phenomenon has been seen in smoothed particle hydrodynamics
simulations (\citealt{Nixon2012a}; \citealt{Nealon2015}), but not as
yet in the grid-based simulations (\citealt{Morales2014};
\citealt{Zhuravlev2014}). Such differential precession could also
potentially give rise to the line energy shifts observed here, via the
same mechanism of illumination of different disk azimuths. More
sophisticated phase-resolved spectral modelling and additional high
quality data in future will allow tomographic mapping of the inner
flow geometry, further informing numerical simulations.

Recently, \cite*{vandenEijnden2016} found evidence in observations of
GRS 1915+105 that some form of differential precession could indeed be
at play (although likely not as extreme as that suggested by
\citealt{Nixon2012}). They show that, in observations displaying an
energy dependent QPO frequency (\citealt{Qu2010}; \citealt{Yan2012}),
the phase of the band with the higher QPO frequency increases faster
than that of the band with the lower QPO frequency. This confirms that
the frequency difference is intrinsic to the source, and can be
explained if, for example, the inner regions of the flow are
precessing slightly faster than the outer regions. Although there is
no energy dependence of the QPO frequency in the observations we
analyse here, H 1743-322 does show an energy dependence of the QPO
frequency for observations with much higher ($\gtrsim 3$ Hz) QPO
frequencies (\citealt{Li2013}).

Our result has implications for black hole spin measurements. Spin
estimates obtained through disk spectral fitting often assume that the
black hole spin aligns with the binary orbit
(e.g. \citealt{Kolehmainen2010}; \citealt{Steiner2012}), which is
incompatible with the precession model. Indeed, recent spectral modelling of
Cygnus X-1 in the soft state implies a $\gtrsim 13^\circ$
misalignment (\citealt{Tomsick2014}). The iron line method provides an independent measure of
inclination, but assumes that the disk extends down to the ISCO,
whereas the precession model assumes an evolving truncation radius. If
the truncation radius really is moving, the shape of the line energy
modulation should change with QPO frequency (\citealt{Ingram2012a}),
which can be tested in future. We also note that the spectral pivoting
and line energy modulation detected here are \textit{non-linear}
changes in spectral shape, which could bias studies of the
time-averaged spectrum. The biases are likely small, but should be
quantified in future with tomographic modeling, since iron line
fitting is sensitive to fairly small spectral distortions. For the
case of active galactic nuclei (AGN), it is unclear if a misaligned
accretion flow is expected in the absence of a binary
partner\footnote{also, it is notoriously difficult to detect a Type C
  QPO analogue due to the very long period expected through mass
  scaling (\citealt{Vaughan2005})}. If there is precession in AGN, it
will not create a bias through non-linear variability, since the
precession timescale would be longer than a typical integration time.

\subsection{Alternative interpretations}

As an alternative to precession, axisymmetric variations in the
accretion geometry can cause changes in the iron line shape. Since the
disk rotational velocity and gravitational redshift both depend on
radius, variation of the disk inner radius throughout a QPO cycle can
cause shifts in the line energy. For the same reasons, changes in the
radial dependence of disk irradiation, perhaps caused by changes in
the vertical extent of the illuminating source, can also drive changes
in the line shape. However, it is very difficult to explain how such
mechanisms could give rise to two maxima in line energy per QPO
cycle. Nonetheless, in future we will explicitly test the precession
model described above against the data presented here, and compare it
to simple axisymmetric alternatives.

\cite{DeMarco2016} recently suggested that the soft lag
  measured in the $0.1-1$ Hz frequency range for the \xmm~data is a
  reverberation lag corresponding to a $\sim 100~R_g$ path
  length. However, this frequency range is dominated by the QPO. Both
  soft and hard lags are routinely observed for QPOs
  (e.g. \citealt{Qu2010}), and the QPO lag is often very different to
  that measured for the broadband noise
  (e.g. \citealt*{Wijnands1999}), and not compatible with a
  reverberation lag (\citealt{Stevens2016}). Moreover, in our
  Fig. \ref{fig:lags} we show that the two QPO harmonics have different
  lags, and so averaging them together has little physical
  meaning. Reverberation lags are still expected to be present of
  course, but will yield much smaller soft lags than the QPO.

\subsection{Anomalous dataset: orbit 1b}

As for the anomalous dataset, orbit 1b, this is puzzling in the
context of any QPO model. The modulations in line energy and flux and
also power-law index are consistent between all the other
datasets. Orbit 1b shows different modulations in all three of these
parameters\footnote{the power-law normalisation is trivially very similar
across all datasets, because QPO phase is defined from the reference
band flux, which tracks the power-law normalisation to a good
approximation, given that the power-law index varies only with small
amplitude.}, as can be seen in Fig. \ref{fig:allmod}. The most
striking is perhaps the large amplitude, and highly statistically
significant ($4\sigma$), modulation in the line flux in orbit 1b. This
may be indicative of a different geometry during orbit 1b. Such a
geometrical change needs to explain the increased iron line flux, the
increased variability in line flux and also the increased width of the
iron line ($0.51$ keV for orbit 1b and $\sim 0.43$ keV for the other
observations). It also needs to be consistent with the only subtle
differences in other diagnostics (such as the full band power spectrum
and the time averaged power-law photon index) and the change needs to
plausibly happen over a $\lesssim 60$ ks time scale. The increased line
flux implies a greater fraction of continuum photons intercept the
disk, which will broaden the line somewhat by increasing the disk
ionisation. The increased variability in line flux suggests that this
fraction \textit{varies} more than for the other datasets. This could
occur if the misalignment angle between the disk and the black hole
spin axes, $\beta$, is somehow larger, since the misalignment between
the disk and inner flow varies between $0$ and $2\beta$ in the
precession model (\citealt*{Veledina2013}; \citealt{Ingram2015a}). This
extra variability in illuminating photons could make line energy
variations due to ionisation changes significantly more important than
for the other datasets. We see in Fig. \ref{fig:allmod} (right, second
panel) that the line flux, and therefore the flux irradiating the
disk, varies by a factor of $\sim 8$ over a QPO cycle for orbit 1b. This means that
the ionisation parameter ($\xi \propto$ illuminating flux) should also
vary by a factor of 8. In Fig. 1 of \cite*{Matt1993}, we see that
varying the ionisation parameter from $\xi \sim 100$ to $\xi \sim 800$
changes the line rest frame energy from $\sim 6.4$ keV to $\sim 6.7$
keV. This modulation in the rest frame line energy should be in phase
with the line flux, and therefore in anti-phase with the line energy
modulation seen in the other datasets. It is unfortunate that
\nustar~was not observing during orbit 1b, otherwise this hypothesis
could have been tested by tracking the reflection hump. Alternatively
(or perhaps additionally), our view may be obstructed by some material
in our line of sight during orbit 1b, which is plausible given the
likely high inclination of H 1743-322. The variable illumination of
the line of sight material will give rise to variable ionisation,
which will imprint itself onto the phase-resolved spectra.

\section{Conclusions}
\label{sec:conclusions}

We find that the iron line centroid energy in H 1743-322 is modulated
on the QPO period with a statistical significance of $3.7\sigma$. We
also find that this modulation has a non-zero second harmonic with a
statistical significance of $3.94\sigma$. Shifts of the line energy
over a QPO cycle are a distinctive prediction of the Lense-Thirring
precession model (\citealt*{Ingram2009}), in which the inner accretion
flow precesses due to the frame dragging effect. Our observation is a typical example of a
Type-C QPO, implying that this class of QPOs in general are driven by
Lense-Thirring precession, and therefore supporting studies that
measure black hole mass and spin using the period of the Type-C QPO in
combination with that of high frequency QPOs (\citealt{Motta2014};
\citealt{Ingram2014}; \citealt*{Fragile2016}). There are still,
however, unanswered questions. We have simply employed
phenomenological modelling to track the iron line here, but more
physical modelling using a self-consistent reflection model will
provide further insight. We will perform this modelling in a future
paper, as well as testing alternative models to precession. The
largest question mark concerns the anomalous dataset, orbit 1b, which
exhibits different parameter modulations to all other datasets (which
all agree with one another).

In future, high quality observations of the same source displaying a
QPO with a higher frequency will provide further insight. The
precession model predicts the disk inner radius to be smaller for
higher QPO frequencies, and therefore we expect the line energy
dependence on QPO phase to have a different shape. Studies such as
this will be greatly enhanced by new instrumentation. Detectors with a
very large collecting area will allow us to perform similar studies
without needing to stack over very long exposures as is necessary
here. Also, X-ray polarimetry will provide an extra dimension,
particularly when combined with phase-resolved spectroscopy
(\citealt{Ingram2015a}). The precession model predicts that the
polarisation angle changes with QPO phase, and that the extrema in
polarisation angle coincide with maxima in the line energy.

\section*{Acknowledgments}

A. I. acknowledges support from the Netherlands Organization for
Scientific Research (NWO) Veni Fellowship, grant number
639.041.437. A. I. thanks Martin Heemskerk for assistance making the
plots in this paper. A. I. acknowledges useful conversations with Dom
Walton, Matteo Bachetti and Brian Grefenstette about \nustar~data
reduction. A. I. acknowledges useful conversations with Andrew King
about disk warping. M. J. M. appreciates support via an STFC Ernest
Rutherford Fellowship. C. D. acknowledges support from the STFC
consolidated grant ST/L00075X/1. D. A. acknowledges support from the
Royal Society. M. A. is an International Research Fellow of the Japan
Society for the Promotion of Science. We thank the anonymous referee
for constructive comments.


\bibliographystyle{/Users/adamingram/Dropbox/bibmaster/mn2e}
\bibliography{/Users/adamingram/Dropbox/bibmaster/biblio}

\appendix

\section{Data visualisation}
\label{sec:datavis}

In order to create the probability maps shown in
Figs. \ref{fig:sepmod} and \ref{fig:allmod}, we run an MCMC in
\textsc{xspec} after finding a best-fit model in the Fourier
domain. \textsc{xspec} uses the \textsc{emcee} algorithm (the MCMC
hammer: \citealt{emcee}). We use the Goodman-Weare algorithm with a
chain length of $3\times 10^5$ steps and $10^3$ walkers. The starting
point of the chain is a randomised realisation of the best-fit
parameters. Visual inspection of the $\chi^2$ implies that the chain
takes $\sim 2\times 10^4$ steps to converge, so we burn $2.5\times
10^4$ steps. For the rest of the chain, the autocorrelation function
of the parameters of interest is centrally peaked, indicating
reasonable convergence. Even so, we note that none of our
significances or error estimates use these chains, we use them purely
for data visualisation.

\begin{figure}
 \includegraphics[height= 7.5cm,width=8.5cm,trim=0.8cm 0.0cm 0.0cm
0.0cm,clip=true]{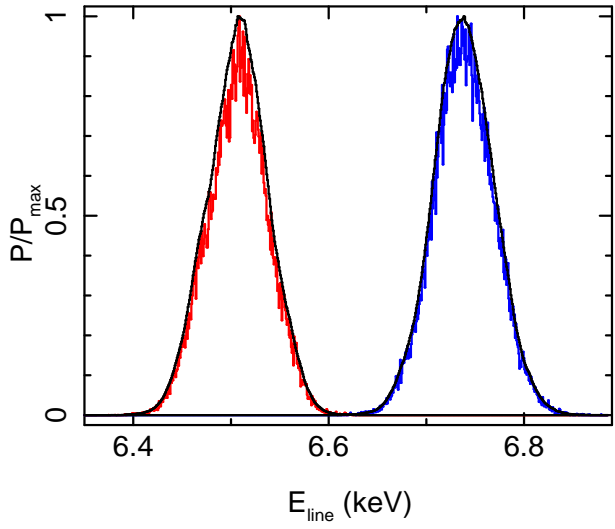}
\vspace{-5mm}
 \caption{Histograms for the line energy for two QPO phases (red:
   phase=7/16 cycles, blue: phase=11/16 cycles) created using an
   MCMC. The black lines are smoothed versions of these histograms
   (see text for details)}
 \label{fig:runbox}
\end{figure}

For the probability maps in Figs. \ref{fig:sepmod} and
\ref{fig:allmod}, we calculate the $E_{line}(\gamma)$ function for
each step of the chain, for 400 values of $\gamma$. That is, for each
step of the chain, we read in the parameters $A_{1E}$, $A_{2E}$,
$\phi_{1E}$, $\phi_{2E}$ and $E_0$ for that step and calculate
$E_{line}(\gamma)$ from equation \ref{eqn:eline}. For each $\gamma$
value, we thus have $2.75\times 10^5$ values of $E_{line}$, which we
bin into an 800 bin histogram. Fig. \ref{fig:runbox} shows these
histograms for the chain corresponding to our joint fit, for two
selected QPO phases (red: phase=7/16 cycles, blue: phase=11/16
cycles). We normalise each histogram to peak at unity. For plotting
purposes, we smooth these histograms by averaging each of the 800 bins
with the $\pm 10$ bins either side. The black lines show the smoothed
versions of the histograms. We use the smoothed versions for our
probability maps.

\label{lastpage}

\end{document}